\documentclass[preprint]{aastex}
\shorttitle{Google Earth Imagery for SDSS Co-added Data}

\shortauthors{Hao and Annis}

\usepackage[usenames]{color}
\usepackage{url}
\usepackage{color}

\begin{document}

\title{Flying across Galaxy Clusters with Google Earth: additional imagery from SDSS co-added data}
\author{Jiangang Hao and James Annis}

\affil{Center for Particle Astrophysics, Fermi National Accelerator Laboratory, Batavia, IL 60510}

\begin{abstract}
Galaxy clusters are spectacular. We provide a Google Earth compatible imagery for the deep co-added images from the Sloan Digital Sky Survey and make it a tool for examing galaxy clusters. More details about how to get it can be found from the following website: \color{blue}\url{https://sites.google.com/site/geclusters/}.
\end{abstract}

\section{Introduction}
Google Earth (in sky mode) provides a highly interactive environment for visualizing the sky. By encoding the galaxy cluster information into a kml/kmz file\footnote{kml/kmz files for Google Earth are just like the html files for web browser.}, one can use Google Earth as a tool for examining galaxy clusters and fly across them freely. 

However, the resolution of the images provided by Google Earth is not very high. This is partially because the major imagery google earth used is from Sloan Digital Sky Survey (SDSS)~\citep{sdss} and the resolutions have been reduced to speed up the web transferring. To have higher resolution images, you need to add your own images in a way that Google Earth can understand.  

The SDSS co-added data are the co-addition of $\sim$100 scans of images from SDSS stripe 82~\citep{annis}. It provides the deepest images based on SDSS and reach as deep as about redshift 1.0. Based on the co-added images, we created color images in a way as described by~\citet{lupton} and convert the color images to Google Earth compatible images using wcs2kml~\citep{brewer}. The images are stored at a public server at Fermi National Accelerator Laboratory and can be accessed by the public. 

To view those images in Google Earth, you need to download a kmz file, which contains the links to the color images, and then open the kmz file with your Google Earth. To meet different needs for resolutions, we provide three kmz files corresponding to low, medium and high resolution images. We recommend the high resolution one as long as you have a broadband Internet connection, though you should choose to download any of them, depending on your own needs and Internet speed.

After you open the downloaded kmz file with Google Earth (in sky mode), it takes about 5 minutes (depending on your Internet connection and the resolution of images you want) to get some initial images loaded. Then, additional images corresponding to the region you are browsing will be loaded automatically. 

So far, you have access to all the co-added images. But you still do not have the galaxy cluster position information to look at. In order to see the galaxy clusters, you need to download another kmz file that tells Google Earth where to find the galaxy clusters in the co-added data region. We provide a kmz file for a few galaxy clusters in the stripe 82 region and you can download and open it with Google Earth. 

\begin{figure*}
\begin{center}
\includegraphics[width=5.5in, height=3.in]{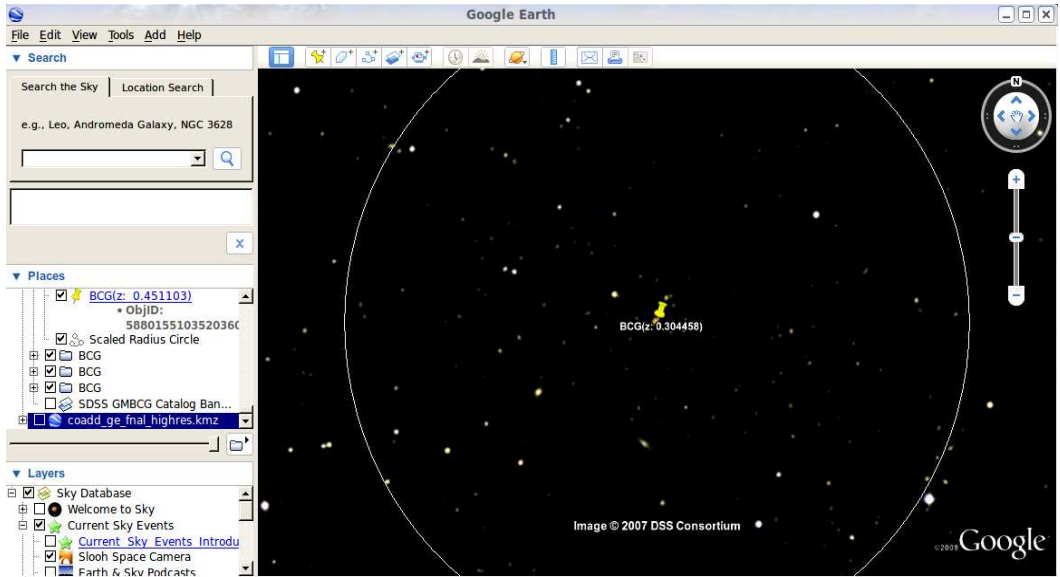}
\caption{Screenshots of a galaxy cluster without co-added SDSS images in Google Earth.}
\label{fig:screen1}
\end{center}
\end{figure*}

\begin{figure*}
\begin{center}
\includegraphics[width=5.5in, height=3.in]{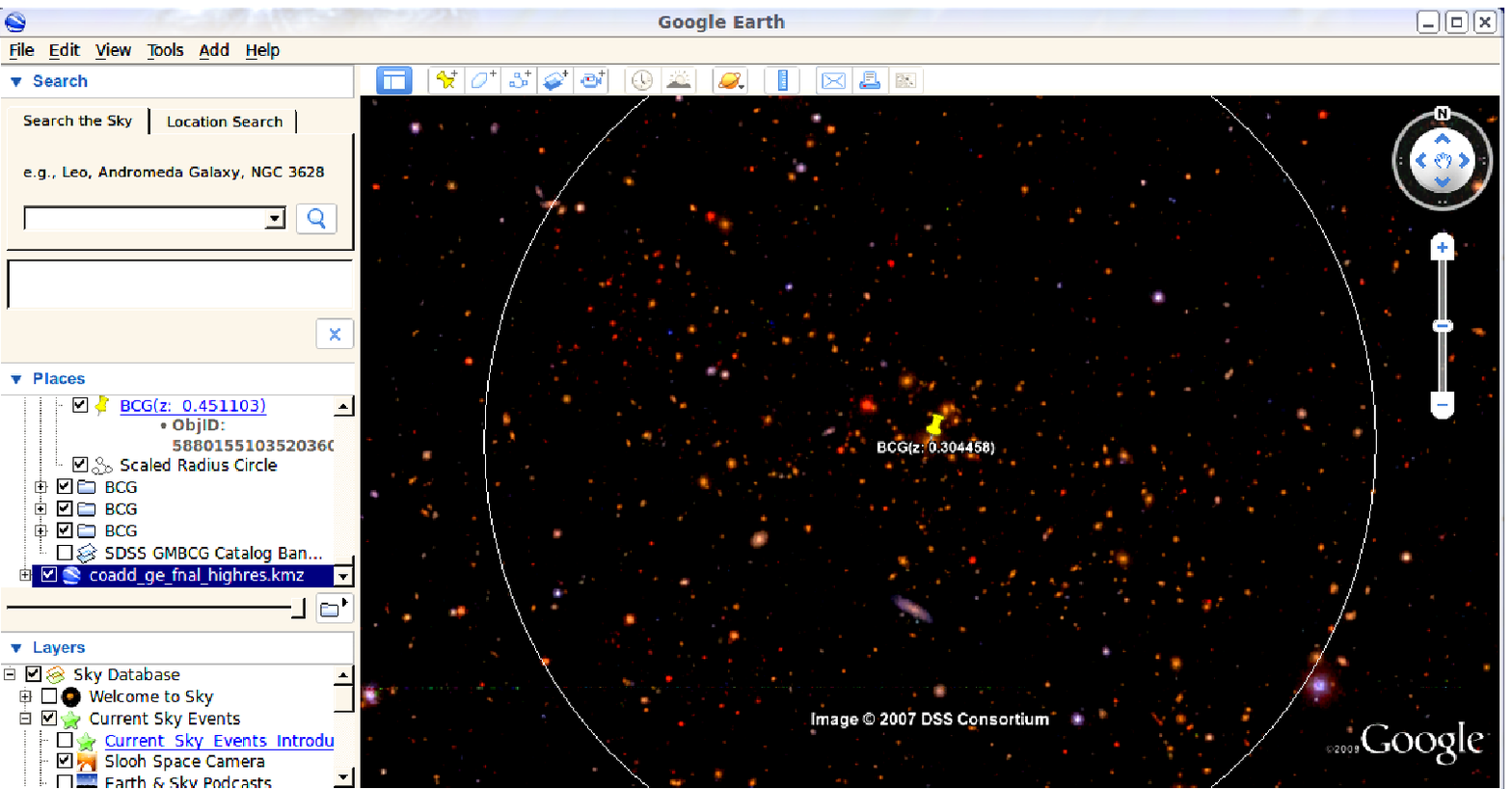}
\caption{Screenshots of a galaxy cluster with co-added SDSS images in Google Earth.}
\label{fig:screen2}
\end{center}
\end{figure*}

In the SDSS co-added region (stripe 82 region), the imagery from Google Earth itself is from the ~\citet{dss}, which is in very poor quality. In Figure\ref{fig:screen1} and Figure\ref{fig:screen2}, we show screenshots of a cluster with and without the new co-added imagery in Google Earth. Much more details have been revealed with the deep images.

\section{More Clusters}
If you want to see more clusters in Google Earth, you can visit our latest GMBCG cluster catalog for SDSS DR7 at: \color{blue}{\url{http://home.fnal.gov/~jghao/gmbcg_sdss_catalog.html}}.

\color{black}
\bibliography{ge}

\section*{Acknowledgments}
Funding for the SDSS and SDSS-II has been provided by the Alfred P.
Sloan Foundation, the Participating Institutions, the National
Science Foundation, the U.S. Department of Energy, the National
Aeronautics and Space Administration, the Japanese Monbukagakusho,
the Max Planck Society, and the Higher Education Funding Council for
England. The SDSS Web Site is http://www.sdss.org/.

The SDSS is managed by the Astrophysical Research Consortium for the
Participating Institutions. The Participating Institutions are the
American Museum of Natural History, Astrophysical Institute Potsdam,
University of Basel, University of Cambridge, Case Western Reserve
University, University of Chicago, Drexel University, Fermilab, the
Institute for Advanced Study, the Japan Participation Group, Johns
Hopkins University, the Joint Institute for Nuclear Astrophysics,
the Kavli Institute for Particle Astrophysics and Cosmology, the
Korean Scientist Group, the Chinese Academy of Sciences (LAMOST),
Los Alamos National Laboratory, the Max-Planck-Institute for
Astronomy (MPIA), the Max-Planck-Institute for Astrophysics (MPA),
New Mexico State University, Ohio State University, University of
Pittsburgh, University of Portsmouth, Princeton University, the
United States Naval Observatory, and the University of Washington.

\end{document}